\documentclass[a4paper,twocolumn,10pt]{quantumarticle}
\pdfoutput=1
\usepackage[utf8]{inputenc}
\usepackage[english]{babel}
\usepackage[T1]{fontenc}
\usepackage{amsmath}
\usepackage{hyperref}
\usepackage{fixltx2e}
\usepackage[numbers,sort&compress]{natbib}

\usepackage[T1]{fontenc}
\usepackage{amssymb}
\usepackage{graphicx}
\usepackage{color}
\usepackage{caption}
\usepackage{hyperref}
\usepackage{amsthm}
\usepackage{amsmath}
\usepackage{braket}
\usepackage{appendix}
\usepackage{subcaption}
\usepackage{comment}
\usepackage{bbold}
\usepackage{multirow}
\usepackage{booktabs}
\usepackage{ulem}
\usepackage[ruled,vlined]{algorithm2e}
\usepackage[dvipsnames]{xcolor}
\usepackage{physics}
\usepackage{enumitem}
\usepackage{float}

\begin{document}

\title{Computing Quantum Resources using Tensor Cross Interpolation}

\author{Sven Benjamin Ko\v{z}i\'c}
\affiliation{%
 Institut Ru\dj er Bo\v{s}kovi\'c, Bijeni\v{c}ka cesta 54, 10000 Zagreb, Croatia.
}%
\orcid{0009-0003-1603-0351}

\author{Gianpaolo Torre}%
 \email{gianpaolo.torre@irb.hr}
\affiliation{%
 Institut Ru\dj er Bo\v{s}kovi\'c, Bijeni\v{c}ka cesta 54, 10000 Zagreb, Croatia.
}%
\orcid{0000-0002-4274-6463}

\begin{abstract}
Quantum information quantifiers are indispensable tools for analyzing strongly correlated systems. 
Consequently, developing efficient and robust numerical methods for their computation is crucial. 
We propose a general procedure based on the family of Tensor Cross Interpolation (TCI) algorithms to address this challenge in a fully general framework, independent of the system or the quantifier under consideration. 
To substantiate our approach, we compute as illustrative examples the non-stabilizerness R\'{e}nyi entropy (SRE) and Relative Entropy of Coherence (REC) for the 1D and 2D ferromagnetic Ising models. 
This method not only demonstrates its versatility, but also provides a generic framework for exploring other quantum information quantifiers in complex systems.
\end{abstract}

\maketitle

\section{Introduction}

The application of quantum information concepts and methods in many-body physics has been highly successful in enhancing our understanding of complex systems.
To provide an example, quantum entanglement has proven crucial in characterizing exotic phases of matter that are not captured by the standard Ginzburg-Landau approach~\cite{Landau1937}, such as topologically ordered phases. In these phases, the global structure of the ground states is associated with the presence of non-local correlations, which can be revealed through their entanglement properties~\cite{RevModPhys.80.517, PhysRevA.71.022315}. A similar picture has also recently been revealed in the so-called topologically frustrated systems, in which frustration is induced by an appropriate choice of boundary conditions, and characterized by the presence of a logarithmic correction of topological origin to the entanglement entropy~\cite{Giampaolo_2019,torre2024interplaylocalnonlocalfrustration,10.21468/SciPostPhysCore.7.3.050, Odavic2023randomunitaries}. Interest in studying quantum resources in many-body systems has also been extended beyond entanglement to include properties such as quantum discord~\cite{PhysRevB.78.224413, Wang_2011}, quantum coherence~\cite{li_quantum_2016}, and, more recently, non-stabilizerness (also known as quantum magic)~\cite{10.21468/SciPostPhys.15.4.131, PRXQuantum.4.040317, Tarabunga2024criticalbehaviorsof, catalano2024magicphasetransitionnonlocal}.

Accurately estimating quantum resources in the thermodynamic limit requires considering sufficiently large systems. This usually makes the calculation of these quantities unfeasible due to the exponentially growing dimension of the corresponding Hilbert space. The issue is twofold: on one hand, there is the need to find an approximate expression for the system's state that captures all the necessary required information. 
On the other hand, given an expression of the state, there is the challenge of developing efficient algorithms for the estimation of a given resource.

The first aspect can be efficiently addressed through the application of Tensor Network (TN) methods~\cite{ORUS2014117,Orus2019,Bridgeman_2017}, which currently represent one of the most powerful tools for the analysis of strongly correlated quantum systems, particularly in low-dimensional regimes.
These methods are based on the possibility of deriving an efficient representation of a quantum many-body state through the amount and structure of its entanglement~\cite{ORUS2014117}. 
As a result, the past two decades have witnessed a steady increase in the number of tensor network (TN) techniques, developed for both static and dynamic settings, specific to different classes of problems (see~\cite{annurev:/content/journals/10.1146/annurev-conmatphys-040721-022705,montangero2018introduction} for a review).

Nevertheless, an efficient representation of the state, while crucial, is not sufficient for the computation of quantum resources. 
Indeed, the calculation of these quantities generally involves evaluating non-linear functions of an exponentially growing number of parameters. 
Consequently, their calculation requires algorithms tailored to the representation of the state and the specific quantifier under consideration. 
One potential solution to this challenge is to employ sampling techniques. 
This method has recently been applied to the computation of quantum magic, both by direct~\cite{PhysRevB.107.035148,Haug2023stabilizerentropies,lami2023quantummagicperfectpauli} and stochastic~\cite{PRXQuantum.4.040317,Tarabunga2024criticalbehaviorsof, PhysRevLett.133.010602} sampling. 
Continuing this trend, a sampling approach has been proposed to evaluate the entanglement entropy in systems of arbitrary dimensions~\cite{PRXQuantum.3.030312}, for which there are currently no other efficient methods. 

In this work we suggest a general method for measuring quantum resources. 
The approach we propose involves expressing the chosen quantifier as a tensor-valued function $F$ to be sampled exploiting the family of Tensor Cross Interpolation (TCI) algorithms~\cite{Oseledets2010, Savostyanov2011,Dolgov2020, fernandez2024}. These classes of algorithms have recently found applications in physics, serving as an efficient, sign-problem-free alternative to Monte Carlo sampling for evaluating high-dimensional integrals in Feynman diagram calculations~\cite{fernandez2022}. Additionally, they have been employed to compute topological invariants~\cite{PhysRevLett.132.056501}, to reconstruct the
exact AKLT state starting from limited number of spin
configurations~\cite{monturiol2025}, and to evaluate overlaps between atomic orbitals~\cite{jolly2024tensorizedorbitalscomputationalchemistry}. TCI deterministically samples $F$ to construct a compressed Matrix Product State (MPS) representation $\tilde{F}$, from which the given measure can be computed straightforwardly. 
The key advantage of employing these algorithms lies in their sampling efficiency: the number of function evaluations to build $\tilde{F}$ is exponentially smaller compared to the growth of the system's dimension. 
Additionally, this sampling procedure is robust; within a specified tolerance, the system reliably converges to the desired results after an appropriate number of function calls.

As a first illustrative example we apply the proposed method for the computation of non-stabilizerness R\'{e}nyi entropy (SRE), that is a well-known witness of the so-called Quantum magic~\cite{PhysRevLett.128.050402}. This property serves as a critical resource for quantum computation, both from a foundational and practical standpoint, as it is an essential component for reaching quantum advantage. 
Indeed, stabilizer states, which are generated via Clifford circuits, can be efficiently simulated on classical computers in polynomial time, regardless of their potentially high degrees of entanglement~\cite{PhysRevA.70.052328}.
The particular structure of SRE has led to the development of several efficient algorithms~\cite{Tarabunga_2024,PhysRevB.107.035148,Haug2023stabilizerentropies,lami2023quantummagicperfectpauli,PRXQuantum.4.040317,Tarabunga2024criticalbehaviorsof,collura2024quantummagicfermionicgaussian,ding2025evaluatingmanybodystabilizerrenyi}. However, our primary aim is to highlight the generality of TCI, that enables the computation of such quantities without the need for specific adaptation to individual measures. To illustrate this, we showcase the procedure for computing the non-stabilizerness of the ground state in the 1D ferromagnetic (FM) Ising model.

Moreover, to prove that adaptability of our approach, we also apply it to the evaluation of quantum coherence, that has been proven to be a powerful tool for investigating foundational problems in quantum mechanics, as it is intrinsically linked to the principle of superposition~\cite{PhysRevLett.113.140401, RevModPhys.91.025001, RevModPhys.89.041003, gao_experimental_2018, liu2024}. 
On the other hand, it is a key quantity in the quantum information and many-body physics, due to its deep connection with the entanglement~\cite{PhysRevLett.115.020403}, and for representing an important tool for the characterization of phase transitions~\cite{li_quantum_2016, PhysRevB.101.115142, doi:10.1073/pnas.1408861112}. 
Recently, it also been proved its connection with many-body localization~\cite{PhysRevA.110.022434} and quantum transport phenomena~\cite{PhysRevB.108.125422}. 
The method we will discuss here will be used in the paper~\cite{Kozic2025} to compute the quantum coherence for the class of topologically frustrated models. 
Here, we instead consider the case of the 2D FM Ising model in a transverse field, applying the same algorithm, thereby demonstrating the generality of our approach. 
We emphasize that, to the best of our knowledge, no other algorithm currently exists for this purpose.

The paper is organized as follows. 
In Section~\ref{TCI_method}, we present the general idea of using TCI for computing quantum correlations. 
In Section~\ref{SRE_section}, as a first application of our method, we compute the non-stabilizerness, as measured by the R\'{e}nyi-$2$ entropy (SRE), for the 1D FM Ising model, while in Section~\ref{REC_section}, we compute the quantum coherence, as measured by the Relative Entropy of Coherence (REC), for the 2D FM Ising model. 
Finally, in Section~\ref{conclusion_and_outlooks}, we discuss the pros and cons of the proposed method and explore further possible generalizations.

\section{Method}\label{TCI_method}
In this section we illustrate the general method for computing quantum resources through sampling, using the family of TCI algorithms. 
The entire process is summarized in Algorithm~\ref{alg:quantum_resources_computation} for clarity.

The quantifiers we consider are scalar-valued functions with tensor arguments~\cite{RevModPhys.91.025001}, defined as $M: \bigcup_\mathcal{H} \mathcal{S}(\mathcal{H}) \to \mathbb{R}{\geq 0}$, where $\mathcal{H}$ represents a Hilbert space and $\mathcal{S}(\mathcal{H})$ denotes the set of density matrices. 
We then assume that these quantifiers take the form $M(\rho) = \sum_{\boldsymbol{\sigma}} F_{\boldsymbol{\sigma}}$, where $F$ is a tensor of degree $\mathcal{L}$ with elements indexed by $\boldsymbol{\sigma} = (\sigma_1, \ldots, \sigma_\mathcal{L})$, with $1 \leq \sigma_l \leq d$, and where $d$ represents the dimension of the individual Hilbert spaces, which we assume to be the same for all components. 
The computation of $M$ thus reduces to evaluating the tensor elements $F_{\boldsymbol{\sigma}}$, which are functions of the input states and, potentially, operators (see Tab.~\ref{table1} for the definition of this function for the SRE and REC measures).

It is important to note that since we consider large systems, the first necessary step in evaluating $M$ is to approximate these states and operators in a form that retains all the essential information. 
This is typically accomplished by expressing them through a tensor network (TN) representation. 
In the following discussion, we assume that both states and operators are represented in this form. However, it is important to note that the sampling procedure we propose is not tied to this specific representation, provided that sampling can be performed.
Furthermore, it is crucial to highlight that obtaining the full tensor $F$ is computationally intractable, as the number of tensor elements increases exponentially with the system size, analogous to the scaling observed for states and operators. 
To overcome this limitation, we then adopt a sampling-based approach exploiting the family of TCI algorithms~\cite{Oseledets2010, Dolgov2020, fernandez2022, fernandez2024}. 
In particular, we focus on the method described in~\cite{fernandez2024}, which is based on the LU decomposition. 
However, it is important to emphasize that the methodology is general and can be applied to any algorithm belonging to this class. 
In the following, we briefly outline the key properties of the algorithm relevant to this study, while directing the reader to the relevant literature for a more comprehensive analysis.

\SetAlgorithmName{Alg.}{Alg.}{List of Algorithms}

\begin{algorithm}[t!]
  \caption{Computation of Quantum Resources}
  \label{alg:quantum_resources_computation}
  
  $f \leftarrow$ Define function according to measure $M$\;
  $x \leftarrow$ Choose the input representation $x$ for the given \\ states/operators\;
  $F \leftarrow$ Construct elements of tensor $F$ using the input $x$\;
  $\tilde{F} \leftarrow$ Apply TCI to yield the interpolative \\ decomposition of $F$\;
  $M(\rho) \leftarrow$ Contract $\tilde{F}$ to compute the quantifier\;

\end{algorithm}

The algorithm we employ takes the input tensor $F$, represented as a function, and generates an approximated tensor in MPS form $ \tilde{F} \approx \left[F_1\right]^{\sigma_1}\left[F_2\right]^{\sigma_2}\cdots\left[F_\mathcal{L}\right]^{\sigma_\mathcal{L}} $. 
The final tensor is obtained by deterministically sampling the initial one, performing approximately $\mathcal{O}(\mathcal{L}d\xi^2)$ calls, where $\xi$ is the bond dimension of the resulting MPS. 
The overall efficiency depends on the number of calls to the initial tensor: the larger the bond dimension $ \xi $, the greater the number of calls required. In turn, $ \xi $ depends both on the specific quantifier and the desired accuracy of the final representation. 
It is important to note that this approach is robust, as it does not result in a suboptimal decomposition of the final tensor. Rather, any decrease in efficiency manifests as a slower convergence rate, which is intrinsically connected to the growth of the bond dimension $\xi$~\cite{fernandez2024}. 
In this way, we are able to circumvent the curse of dimensionality, while preserving, within a specified tolerance, all the crucial information necessary for the efficient computation of the given quantum resource. We can finally perform a fast contraction of the new $\tilde{F}$ tensor in the MPS form to obtain the value of the desired measure. In the next two sections we apply the proposed algorithm to the computation of the SRE and the REC respectively. \\

\begin{table*}[!t]
\centering
\begin{tabular}{@{}cccc@{}}
\toprule
\textbf{Measure} & \textbf{Function} & \textbf{Input $x$} & \textbf{Complexity} \\ 
\midrule
SRE              & $f(x)=x^{4}$             & \raisebox{- 0.44 \totalheight}{\includegraphics[width=0.3\textwidth]{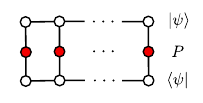}} & $\mathcal{O}(\mathcal{L}\chi^3)$  \\
\hline
\\[-1cm]
REC            & $f(x) = x^{2}\log_{2}(x^{2})$           & \raisebox{-0.44 \totalheight}{\includegraphics[width=0.3\textwidth]{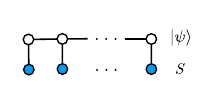}} & $\mathcal{O}(\mathcal{L}\chi^2)$  \\
\\[-0.8cm]
\bottomrule
\end{tabular}
\caption{Sampled function for the SRE and REC}
\label{table1}
\end{table*}

\section{Non-stabilizerness}\label{SRE_section}
The Stabilizer R\'{e}nyi-$2$ Entropy (SRE) has been recently introduced as a measure of non-stabilizerness~\cite{PhysRevLett.128.050402}. For a pure state $\ket{\psi}$ defined on the space of $L$ qubits, it is defined as 
\begin{equation}\label{magic}
    \mathcal{M}_2(\ket{\psi}) \equiv  - \log_2 \left(\dfrac{1}{2^L}\sum_P \ev{P}{\psi}^4\right),
\end{equation}
where the sum runs on all the Pauli strings $P=\bigotimes_{j=1}^L P_j$, with $P_j \in \left\lbrace \mathbb{1}, X, Y, Z\right\rbrace$, and $X,Y,Z$ are the Pauli matrices. The computation of the SRE given in Eq.~\eqref{magic} has recently garnered significant attention, with various efficient approaches now available~\cite{Tarabunga_2024,PhysRevB.107.035148,Haug2023stabilizerentropies,lami2023quantummagicperfectpauli,PRXQuantum.4.040317,Tarabunga2024criticalbehaviorsof,collura2024quantummagicfermionicgaussian,ding2025evaluatingmanybodystabilizerrenyi}.
In this context, we emphasize the strength of our method, which lies in its generality: the algorithm remains consistent across various systems, as it is fundamentally rooted in the definition itself. We can compute the SRE by rewriting Eq.~\eqref{magic} as $M_2(\ket{\psi}) = - \log_2 \left( \frac{1}{2^L} \sum_P f(P) \right)$, where $f(P)=\ev{P}{\psi}^4$ is the function to be sampled. Since we are working with the MPS representation of the input state, the computation of $f(P)$ in terms of TN contractions is described in Table~\ref{table1}.

The computational cost of the algorithm arises from the number of calls to the function $f$, which is determined by the bond dimension of the resulting MPS $\tilde{F}$, as well as the computational cost connected to each individual call to $f$. The contraction we consider (see Tab.\ref{table1}) is known to scale as $O(\mathcal{L}\chi^3)$~\cite{Haug2023stabilizerentropies,lami2023quantummagicperfectpauli,PhysRevB.85.165146}, being $\chi$ the bond dimension of the input MPS. 
Moreover, considering that the number of function calls scales with the bond dimension of the final MPS as $\mathcal{O}(\mathcal{L}\xi^2)$, the overall computational complexity is given by $\mathcal{O}(2\mathcal{L}^2\xi^2\chi^3)$.

\begin{figure}[t!]
    \centering
    \begin{subfigure}[t]{0.48\textwidth}
        \centering
        \includegraphics[width=1.0\textwidth]{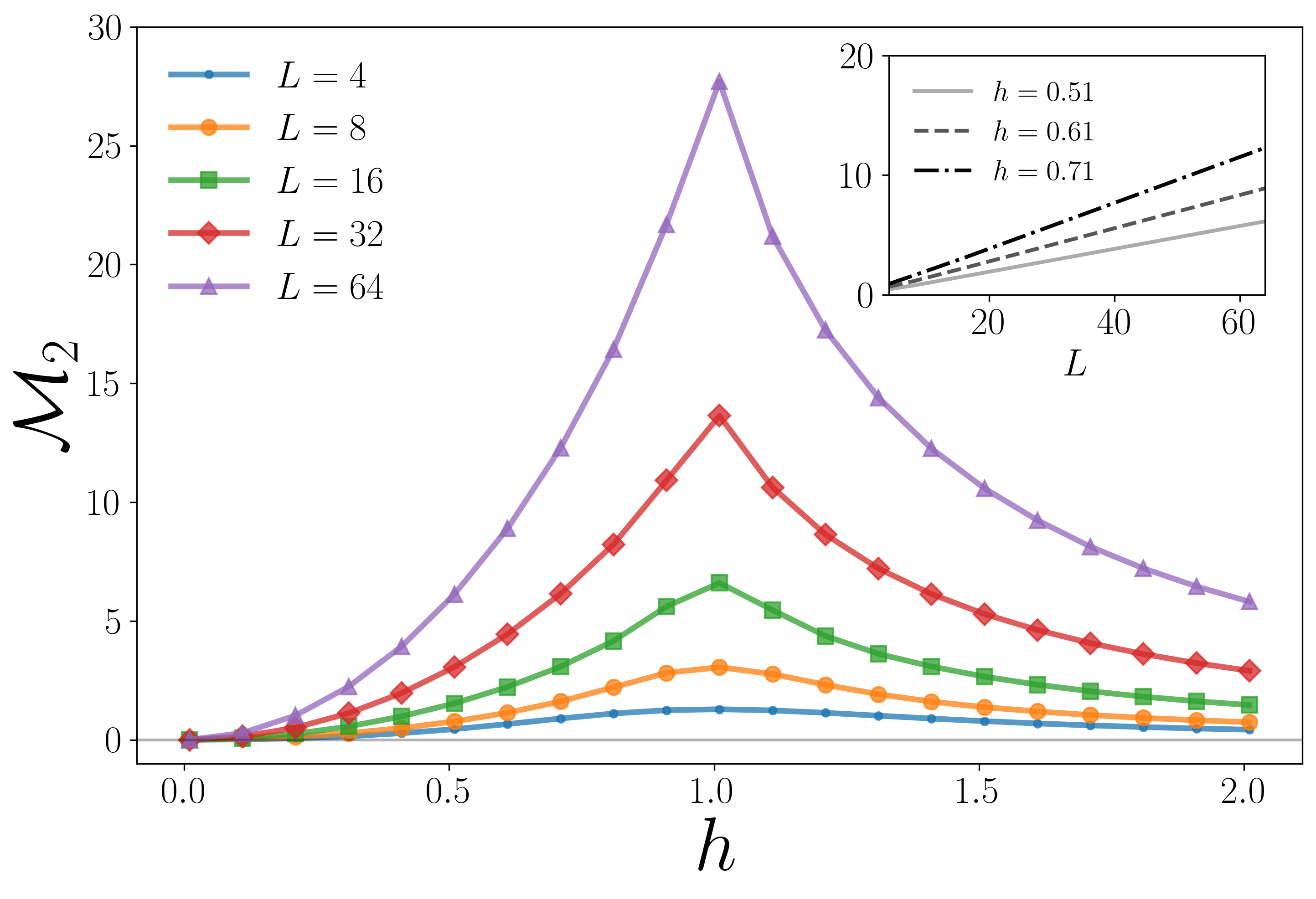}
    \end{subfigure}
    \caption{The Stabilizer Rényi entropy Eq.~\eqref{magic} for the 1D ferromagnetic (FM) Ising model Eq.~\eqref{1D_FM_Ising} is presented as a function of the magnetic field for various system sizes. The maximum bond dimension of the input state was constrained to $\chi \leq 50$, while the maximum bond dimension of the TCI MPS was set to $\xi = 80$.}
    \label{Qmagic_2D_FM_Ising}
\end{figure}

To illustrate the method, we evaluate the SRE for the ground state of the 1D FM transverse-field Ising chain.
\begin{equation}\label{1D_FM_Ising}
    \mathcal{H}=-\sum_{j=1}^{L}Z_j Z_{j+1}-h\sum_{j=1}^{L}X_j,
\end{equation}
where $X_j, Z_j, \forall j=1,\ldots,L$ are the $X,Z$ Pauli operators acting on the $j$-th spin and $h$ is the magnetic field. We assumed periodic boundary conditions, hence $Z_{L+1} = Z_L$. The SRE behavior for this system is well-established from previous work~\cite{PhysRevA.106.042426,10.21468/SciPostPhys.15.4.131}, and therefore we refer to the literature for a more detailed analysis. We would like to emphasize that in this work, we were able to reach a system size of $L=64$ spins without relying on any specific implementation of the algorithm, but by directly applying the definition Eq.~\eqref{magic}.

The results are reported in Fig.~\ref{Qmagic_2D_FM_Ising}. In the limit $h \rightarrow 0^+$, we found that the SRE is zero. Indeed, in this case, the system ground state is given by the equal-weight superposition $\ket{g} = \frac{1}{\sqrt{2}} (\ket{\uparrow}^{\otimes L} + \ket{\downarrow}^{\otimes L})$, with $\ket{\uparrow}$ ($\ket{\downarrow}$) being the eigenstates of the $Z$ operator. This is a GHZ state, which is known to have zero magic~\cite{PhysRevA.106.042426, 10.21468/SciPostPhys.15.4.131}. Moving into the FM phase, we observe an increase in the SRE as the magnetic field grows. The inset highlights that this increase follows a linear trend, indicating that the SRE remains localized. This behavior is linked to the locality of the interactions and the fact that the system remains gapped~\cite{PhysRevA.106.042426}. \\

\section{Relative Entropy of Coherence}\label{REC_section}
Quantum coherence can be quantified using several different measures~\cite{RevModPhys.89.041003}. In this work, we consider the Relative Entropy of Coherence (REC), which has the advantage of being expressed in a simple form in terms of the density matrix elements and its eigenvalues. Furthermore, its expression becomes even simpler in the case of pure states.
The REC for a density matrix $\rho$ is defined as $\mathcal{C}(\rho) \equiv \mathcal{S}(\rho_{\textrm{diag}}) - \mathcal{S}(\rho)$, where $\mathcal{S}(\rho)$ denotes the von Neumann entropy, and $\rho_{\textrm{diag}}$ is the diagonal matrix obtained by setting the off-diagonal elements of $\rho$ to zero. Since we are considering pure states, $\mathcal{S}(\rho) = 0$, and the expression simplifies to
\begin{equation}\label{QC}
    \mathcal{C}(|\psi\rangle) = - \sum_{i=1}^{2^L} c_i^2 \log_2 c_i^2.
\end{equation}
Here, $c_i$ are the coefficients of the state $\ket{\psi} = \sum_{i=1}^{2^N} c_i \ket{i}$ in the chosen basis $\{\ket{i}\}_{i=1}^{2^L}$. Eq.~\eqref{QC} represents a notable example of the use of our approach. Indeed, despite its simplicity, the number of coefficients to be considered for a given approximation grows exponentially with the system size. The use of the TCI algorithm for an intelligent sampling of Eq,~\eqref{QC} then represents, as far as we know, the most effective approach.

We can measure the REC applying the illustrated method by rewriting Eq.~\eqref{QC} as $\mathcal{C}(|\psi\rangle) = - \sum_{S}f(S)$, where the function is defined as $f(x) = x^{2}\log_2(x^2)$, with $x\equiv\braket{S}{\psi}$, $\ket{S}=\left\{\ket{\uparrow},\ket{\downarrow}\right\}^{\otimes L}$. The function $f$ is simply the operation of retrieving a component from the ground state MPS (see Tab.~\ref{table1}).

The computational cost of the algorithm is straightforwardly determined as $\mathcal{O}(2\mathcal{L}^2\xi^2\chi^2)$, noting that the cost of retrieving a component from the input MPS is $\mathcal{O}(\mathcal{L}\chi^2)$, where $\chi$ is the bond dimension of the input MPS.

This approach will be employed in~\cite{Kozic2025} to compute REC for the class of topologically frustrated spin chains. To demonstrate its versatility, we apply it here to the 2D FM Ising model, described by the Hamiltonian:
\begin{equation}\label{2D_Ising}
    \mathcal{H} = - \sum_{\langle i,j \rangle} Z_i Z_j - h \sum_i X_i,
\end{equation}
where the first summation runs over nearest-neighbor pairs, and periodic boundary conditions (PBCs) are assumed. This model is well-studied in the literature~\cite{suzuki2012quantum,sachdev2011,McCoyWu+1973}, and it is known to undergo a phase transition at $h_c \approx 3.044$, transitioning from a ferromagnetic to a paramagnetic phase as $h$ increases~\cite{PhysRevB.57.8494,PhysRevE.66.066110}. In our analysis, we focus on the ferromagnetic regime, $h < h_c$, where the Ising interaction dominates over the external magnetic field.

\begin{figure}[t!]
    \centering
    \begin{subfigure}[t]{0.49\textwidth}
        \centering
        \includegraphics[width=1.0\textwidth]{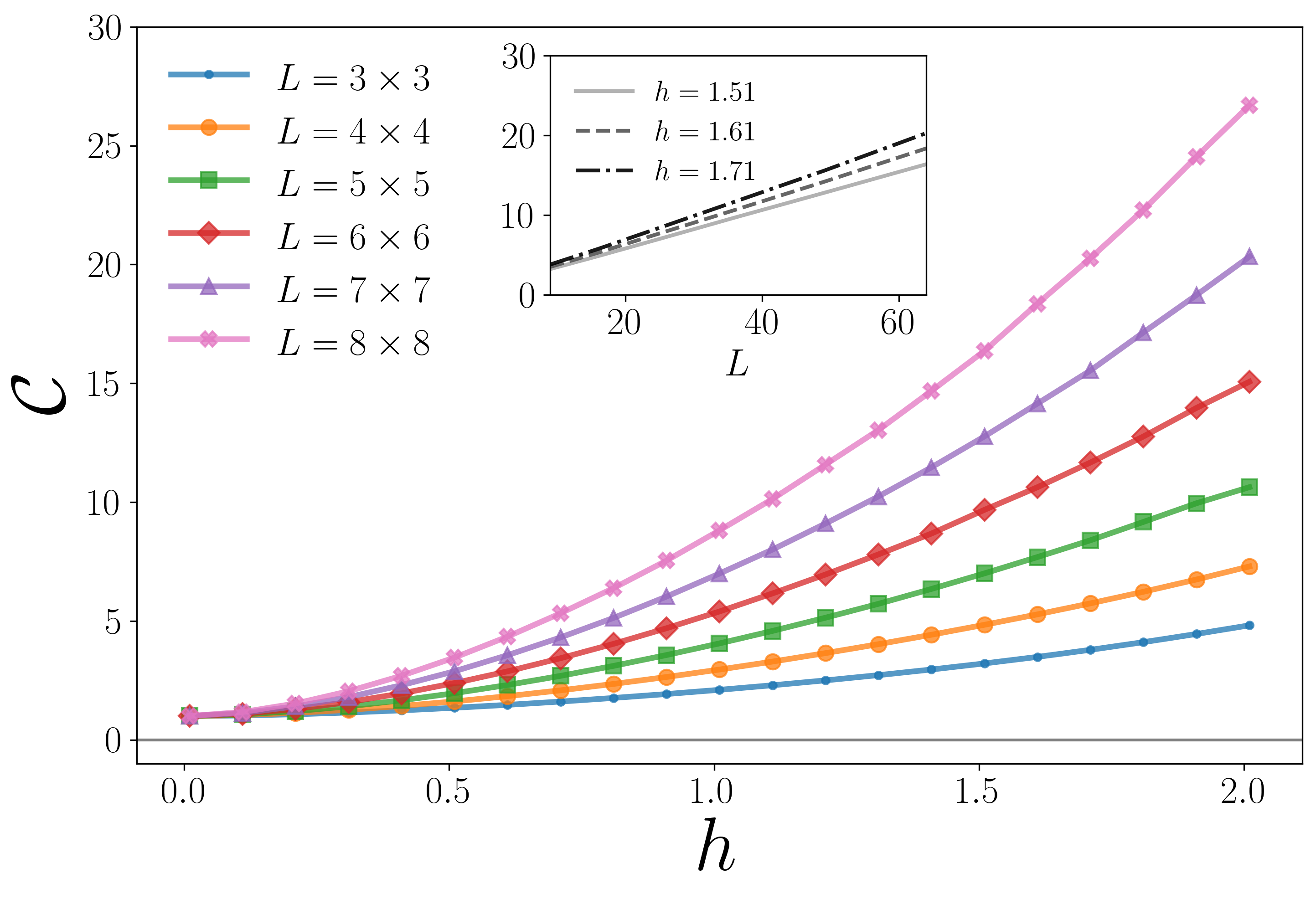}
    \end{subfigure}
    \caption{Quantum coherence Eq.~\eqref{QC} for the 2D FM Ising model Eq.~\eqref{2D_Ising} as a function of the magnetic field for different system size. The maximum bond dimension of the input state was constrained to $\chi \leq 50$, while the maximum bond dimension of the TCI MPS was set to $\xi = 40$.}
    \label{QC_2D_FM_Ising}
\end{figure}

The results are summarized in Fig.~\ref{QC_2D_FM_Ising}, where the REC is plotted as a function of the magnetic field $h$ for different system sizes. We first observe that, in the limit $h \rightarrow 0^+$, $\mathcal{C}(\rho)$ becomes independent of $L$ and equals 1, as the ground state can be written as the equal-weight superposition $\ket{g} = \frac{1}{\sqrt{2}}(\ket{\uparrow}^{\otimes L} + \ket{\downarrow}^{\otimes L})$, with $\ket{\uparrow}$ ($\ket{\downarrow}$) being the eigenstates of the $Z$ operator with eigenvalues $+1$ ($-1$). As we move into the phase, the QC shows a dependence on the magnetic field. For the 1D FM Ising chain this dependence follows a volume law~\cite{Kozic2025}, meaning REC increases linearly with the system length. The same volume law dependence is observed in the 2D case: as shown in the inset, the REC scales linearly with the system size. \\

\section{Conclusions and outlooks}\label{conclusion_and_outlooks}
We proposed a general approach for computing quantum resources, based on the sampling of the corresponding quantifier. 
This sampling is performed efficiently by using the family of TCI algorithms. 
These algorithms enable computations to be carried out with a power-law scaling of the number of operations, i.e., with an exponentially reduced number of operations relative to the number of parameters defining the states and operators involved. 
In this way, we can circumvent the exponential growth of the Hilbert space while preserving all relevant information.

We have examined two paradigmatic examples of quantum resources: non-stabilizerness, quantified by the stabilizer Rényi-$2$ entropy, and quantum coherence, measured using the Relative Entropy of Coherence. 
For the first case, we compute the SRE for the ground state of the 1D ferromagnetic Ising model, confirming the results found in~\cite{PhysRevA.106.042426,10.21468/SciPostPhys.15.4.131}. We would like to emphasize again that, although efficient algorithms exist~\cite{PhysRevB.107.035148,Haug2023stabilizerentropies,lami2023quantummagicperfectpauli,PRXQuantum.4.040317,Tarabunga2024criticalbehaviorsof,collura2024quantummagicfermionicgaussian}, we were able to reach a dimension of up to $L=64$ spins just by directly applying the definition of the measure. We would also like to point out that this method is not limited specifically just to the non-stabilizerness R\'{e}nyi-$2$ entropy but can also be used, for example, in the estimation of non-local magic defined in~\cite{Cao2024}. In the second case we calculate the REC for the 2D ferromagnetic Ising model. 
It is important to note that, to the best of our knowledge, no other efficient algorithm exists for this purpose. We thus demonstrate that the same approach enables computations well beyond the limits of the exact diagonalization method, while also being adaptable and general enough to be applied to a wide range of models and measures. Furthermore, we wish to emphasize that the ability to access non-linear functions of TNs opens new possibilities in bridging the gap between traditional machine learning techniques and tensor networks~\cite{monturiol2025}.

The overall efficiency of the approach relies on two main factors: the computational cost required to evaluate the function $f$ and the number of function calls performed by the TCI algorithm. 
Regarding the first factor, selecting the optimal representations for states and operators tailored to the problem at hand can improve the algorithm's efficiency. 
In this work, we use the TN MPS representation for the ground state in both the 1D and 2D FM Ising models. 
While a TTN representation would be more suitable for the 2D case, this alternative approach and a more substantial analysis is left for future explorations. As for the second factor, the number of calls to the function $f$ depends on the bond dimension of the output MPS, scaling as $\mathcal{O}(\mathcal{L}\xi^2)$. 
It is worth noting that the standard implementation of the TCI algorithm produces an MPS tensor as output, which may not always be optimal for certain problems, such as 2D models. Exploring algorithms that yield alternative tensor network structures as output can address this issue. 
For instance, a recent study~\cite{Tindall2024} suggests promising directions in this regard, and we plan to investigate these alternatives in future works.\\

\section{Acknowledgments}
SBK and GT express their gratitude to Simone Montangero, Nora Reinić, and Daniel Jaschke for their support in the extensive field of tensor networks, during their visit at the University of Padova. They also acknowledge Salvatore Marco Giampaolo and Fabio Franchini for encouragement and for carefully reviewing the manuscript. The code for calculating the ground state of the 1D and 2D ferromagnetic (FM) Ising models has been implemented using the ITensor library~\cite{10.21468/SciPostPhysCodeb.4,10.21468/SciPostPhysCodeb.4-r0.3}. The implementation of the TCI algorithm employed in this work is provided by the \texttt{TensorCrossInterpolation.jl} library~\cite{fernandez2024,TensorCrossInterpolation.jl}. All the codes are provided in the \texttt{QuantumResourcesTCI} repository~\cite{kožić_torre_2025}.

\bibliography{references}
\bibliographystyle{quantum}

\end{document}